\documentclass{elsart}

\usepackage{amssymb}
\usepackage{ae,aecompl}
\usepackage{epsfig}
\usepackage{graphics}
\usepackage{graphicx}
\usepackage{amsmath}
\usepackage{pifont}
\usepackage{psfrag}
\begin{document}

\begin{frontmatter}



\title{Increasing market efficiency: Evolution of cross-correlations
of stock returns}


\author[bme]{Bence T\'oth},
\ead{bence@maxwell.phy.bme.hu}
\author[bme,lce]{J\'anos Kert\'esz}

\address[bme]{Department of Theoretical Physics, Budapest University
of Technology and Economics, Budafoki út 8, H-1111 Budapest, Hungary}
\address[lce]{Laboratory of Computational Engineering, Helsinki
University of Technology, P.O.Box 9400, FIN-02015 HUT, Finland}

\begin{abstract}
We analyse the temporal changes in the cross correlations of returns
on the New York Stock Exchange.  We show that lead-lag relationships
between daily returns of stocks vanished in less than twenty years. We
have found that even for high frequency data the asymmetry of time
dependent cross-correlation functions has a decreasing tendency, the
position of their peaks are shifted towards the origin while these peaks
become sharper and higher, resulting in a diminution of the Epps
effect. All these findings indicate that the market becomes
increasingly efficient.
 \end{abstract}

\end{frontmatter}

\section{Introduction}
\label{sec:intro}
Correlation functions are basic tools in statistical physics.
Equal-time cross-correlations are related to thermodynamic second
derivatives of potentials, i.e. to generalised
susceptibilities. Time-dependent cross-correlation functions are
important for determining transport coefficients through the
fluctuation-dissipation theorem. The Onsager relations of the
transport coefficients for crossed effects have their roots in the
symmetry properties of time-dependent cross-correlations. These
properties are due to the detailed balance, which on turn is the
consequence of microscopic reversibility.

It is natural that in econophysics
\cite{stanley.mantegna,bouchaud.potters} much of the studies
concentrates on the correlations between time series obtained from the
stock market.  The study of correlations of returns are crucial for
understanding the mechanisms and the structure of markets. It is
enough to mention here the the central role played by correlations in
the classical theory of portfolio optimisation see,
e.g. \cite{bouchaud.potters}, or the more recent results on their use
in uncovering market taxonomy \cite{mantegna.mst1} by using some
clustering technique, e.g., that of a minimum spanning tree.

In spite of its noisiness \cite{PARIS.randommatrix,BU}, for daily
returns the correlation matrix contains much useful information
\cite{mantegna.mst1,jpo.clustering,Pafka.Kondor}.  Studying equal time
correlations among stock price changes during short intervals Epps
found an interesting phenomenon \cite{epps}: Correlations decreased
with the decrease of the length of the interval for which the price
changes were calculated.  Changes in prices (and thus logarithmic
returns) for longer intervals are merely non-overlapping sums of
price changes for shorter periods, thus the possible causes of this
effect can be non-stationarity, lagged autocorrelations and lagged
cross-correlations of stock returns.  Later other researchers studying
correlations on high frequency data reached the same conclusion,
correlations decrease as the window width, in which the returns are
computed, decrease
\cite{kullmann,iori.epps,reno.epps,drozdz.epps,quantitative.finance2}.

Of course, in contrast to physical systems, in economic systems
detailed balance or time reversal symmetry is not present
\cite{kullmann2,kertesz.pricedrops}. Nevertheless, the study of
time-dependent cross-correlations between stock returns is of great
interest. Time-dependent cross-correlation functions between pairs of
companies may give us information not only about the relation of the two
companies but also on the internal structure of the market and on the
driving mechanisms.

In 1990, Lo et al. \cite{lo} studied an arbitrage possibility, called
contrarian strategy. They showed that the cause of the high
profitability of these contrarian strategies is not only the pattern
of overreaction on the market (negative autocorrelation of returns in
time) but also lead-lag effects (positive cross-correlations of
returns in time). They studied weekly return data of stocks from 1962
to 1987.  They divided the almost 5000 studied stocks into 5 quintiles
of the same size, on the basis of their market values at the end of
the sample period (the \(1^{st}\) quintile containing the largest
stocks and the \(5^{th}\) containing the smallest ones). They created
an equal-weighted return index for each of the quintiles and
calculated the time-dependent correlation matrices with a time shift
of one week, two weeks, three weeks and four weeks. Studying these
matrices they found a distinct lead-lag relation based on size. The
returns of smaller stocks were correlated with past returns of larger
stocks but not vice versa, resulting in a kind of forecastability
of price changes.  

Forecastability contradicts the efficient market hypothesis. We know
that the speed of information flow increases in time which is expected
to act in favour of efficiency. Indeed, Kullmann et al. \cite{kullmann}
needed high frequency data of 1997 and 1998 to find lead-lag
effects. They studied time-dependent cross-correlations of stocks from
the New York Stock Exchange (NYSE) and found relevant lead-lag effects
on the minute scale only. They fixed thresholds to determine the
relevant correlations and also found that larger stocks were more
likely to pull smaller ones but there were some exceptions.

The time evolution of equal time correlations has been studied and for
daily data considerable robustness was observed
\cite{jpo.taxonomy,mantegna.degreestability}, though at crash periods
a phase transition-like change appeared \cite{jpo.blackmonday}.  In
the present paper we study the evolution of time-dependent
cross-correlation functions and equal time cross-correlation
coefficients between logarithmic returns of stocks from the point of
view of market efficiency.

The paper is built up as follows. In Section 2 we first introduce the
data sets, and the way we processed them in order to carry out the
computations. After that we describe the methodology of our
computations. Section 3 contains the results. The paper terminates
with a discussion in Section 4.
\section{Methodology}
\label{methodology}
\subsection{Data and data processing}
\label{data.and.data.processing}
We used two databases for our analysis.
In the study of changes of cross-correlations in high frequency data
we used tick-by-tick data (containing every trade for each stock)
obtained from the Trade and Quote (TAQ) Database of New York Stock
Exchange (NYSE) for the period of 4.1.1993 to 31.12.2003. We confined
ourselves to analysing the 190 most frequently traded stocks in this
period.  The high frequency data obtained from the TAQ Database was
raw data. It contained many technical information of the NYSE stocks:
prices of trades, bid prices, ask prices, volume of trades, etc. There
were separate files containing the data for dividends of the
stocks. To be able to carry out computations, first we had to make the
dividend adjustment by using the dividend files. Next, we created the
logarithmic return time series from the price time series.  To avoid
the problems occurring from splits in the prices of stocks, which
cause large logarithmic return values in the time series, we applied a
filtering procedure. In high-frequency data, we omitted returns larger
than 5\% of the current price of the stock. This retains all
logarithmic returns caused by simple changes in prices but excludes
splits which are usually half or one third of the price.

In the study of changes of time-dependent cross-correlation functions
on daily scale we used daily data of stocks, obtained from Yahoo
financial web page \cite{yahoo.financial} for the period of 4.1.1982
to 29.12.2000. These contained the daily closing prices of 116 large
stocks from NYSE. These data were needed because we went back with our
analysis to times preceding 1993.  The daily prices were already
dividend adjusted and split adjusted so we did not need to carry out a
filtering on them. We just created their logarithmic return time
series for the computations.

\subsection{Equal-time and time-dependent correlations}
We studied the cross-correlations of stock returns in function of time
shift between the pairs' return time series.  In the computations we
used logarithmic returns of stocks:
\begin{equation}
r_{\Delta t}(t)=\ln\frac{p(t)}{p(t-\Delta t)},
\end{equation}
where \textit{p(t)} stands for the price of the stock at time \textit{t}.
The equal-time
correlation coefficient  \(\rho_{\Delta t}^{A,B}\) of stocks \textit{A}
and \textit{B} is defined by
\begin{equation}
\label{eq:corrcoefficient}
\rho_{\Delta t}^{A,B}=\frac{\langle r_{\Delta t}^{A}(t)r_{\Delta
    t}^{B}(t)\rangle -\langle r_{\Delta t}^{A}(t)\rangle\langle
    r_{\Delta t}^{B}(t)\rangle}{\sigma_{A}\sigma_{B}}.
\end{equation}
The time-dependent correlation function \(C_{\Delta t}^{A,B}(\tau)\)
between stocks \textit{A} and \textit{B} is defined by
\begin{equation}
\label{eq:timedepcorr}
C_{\Delta t}^{A,B}(\tau)=\frac{\langle r_{\Delta t}^{A}(t)r_{\Delta
    t}^{B}(t+\tau)\rangle -\langle r_{\Delta t}^{A}(t)\rangle\langle
  r_{\Delta t}^{B}(t+\tau)\rangle}{\sigma_{A}\sigma_{B}}.
\end{equation}
The notion \(\langle \cdots\rangle\) stands for the time average over the
considered period and \(\sigma^{2}\) is the variance of the return series:
\begin{equation}
\label{eq:sigma}
\sigma^{2}=\langle \lbrack r_{\Delta t}(t)-\langle r_{\Delta
t}(t)\rangle\rbrack^{2}\rangle.
\end{equation}
Obviously the equal-time correlation coefficient can be obtained by
setting \(\tau = 0\) in \eqref{eq:timedepcorr}.

In order to avoid major return values in high frequency data, caused
by the difference in opening prices and previous days' closing prices
(which doesn't give us information about the average behaviour of
returns), we took the average in two steps. First we carried out the
average over the intra-day periods and then over the independent days.
In the analysis of the daily data the average was taken in one step
over the time period examined.

We follow and briefly summarise the method used in \cite{kullmann} in
determining the value of \(\Delta t\), the window width in which
logarithmic returns are computed. Since in high-frequency data the
smallest interval between two trades is one second, \(\Delta t = 1\)
second seems to be at first sight a natural choice. Nevertheless,
choosing such a short window when computing the logarithmic returns
would result in a too noisy correlation function. In order to avoid
this problem we chose a wider window for the computation of the
logarithmic returns and averaged the correlations over the starting
points of the returns.  In this way the average in
\eqref{eq:timedepcorr} means:
\begin{equation}
\langle r_{\Delta t}^{A}(t)r_{\Delta t}^{B}(t+\tau)\rangle=\frac{1}{T}
\sum_{t_{0}=0}^{\Delta t-1}\sum_{k=1}^{T/\Delta t}r_{\Delta
t}^{A}(t_{0}+k\Delta t)r_{\Delta t}^{B}(t_{0}+k\Delta t+\tau),
\end{equation}
where the first sum runs over the starting points of the returns and
the second one runs over the \(\Delta t\) wide windows of the returns.
Choosing this window wider can be understood as an averaging or
smoothing of the correlation function.  On the other hand, \(\Delta
t\) should not be chosen too large since this would cause the maximum
of the function to be indistinct.  In accordance with market processes
and papers in the subject \cite{kullmann,kullmann2}, we chose \(\Delta
t = 100\) seconds\footnote{In \cite{kullmann} a model calculation was
presented to demonstrate the method. In that paper in Figure 1 the
value of \(\sigma\) is erroneously given as \(1000\) instead of
\(\sqrt{1000}\).}. When studying daily logarithmic returns \(\Delta
t\) is obviously the smallest time difference in the time series,
i.e. 1 day.

\section{Results}
\label{results}

\subsection{Time-dependent correlations on daily scale}
\label{time-dependent.correlations.on.daily.scale}
\begin{figure}
\begin{center}
\psfrag{ev}[][][1.3][0]{year} 
\psfrag{korr}[][][1.3][0]{\(\overline{C}\)}
\includegraphics[height=100mm]{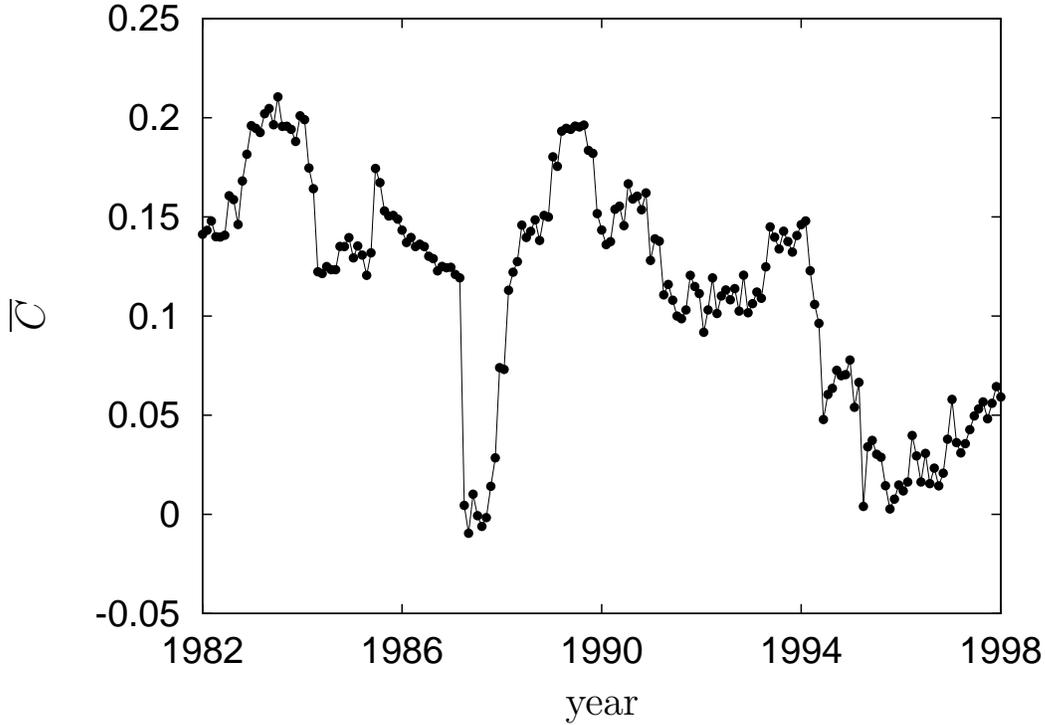}
\caption{The average correlation between highly capitalised and less
capitalised stocks with time shift of one day. The trend is that the
pulling of weaker stocks by stronger stocks diminishes on the daily
scale. The dip is due to the crash in 1987.}
\label{corr.yearly}
\end{center}
\end{figure}

As mentioned before, Lo et al. \cite{lo} found relevant correlations
between weekly returns of stocks data from the sixties, seventies and
eighties. They found that the returns of highly capitalised stocks
pull those of lower capitalised ones much stronger than vice versa. We
studied these results from the point of view of market efficiency.  We
computed the time-dependent correlations on logarithmic returns
obtained from daily closing prices of stocks for the period of 1982 to
2000. On weekly returns we did not find any relevant
correlations. Therefore we carried out our computations on time series
of daily logarithmic returns.  To be able to investigate the average
dynamics of the pulling effect (influence) of highly capitalised
stocks on weaker ones, we had to introduce categories of stocks.  We
divided the stocks in two equal size groups with respect to their
market capitalisation data of \(31^{st}\) December, 1999
\cite{nyse.mc}. We computed an average logarithmic return time-series
for each group.  The average was simply taken with equal weights as in
A. Lo et al. \cite{lo}. We computed the time-dependent correlations
between the average logarithmic return time-series of the group of
larger stocks and that of the group of smaller stocks. Formally this
means that we used \(\Delta t = 1\) day and \(\tau = 1\) day in
\eqref{eq:timedepcorr} in the calculation of \(C_{\Delta t = 1
day}^{big, small}(\tau = 1\) day). We carried out our computations in
2 year wide windows. We shifted the window in every step with 1/10
year (25 trading days).  On Figure \ref{corr.yearly} we can see the
correlation coefficient changing through the years. One can see the
trend of the curve: the correlation is diminishing. From about
\textit{0.15--0.2} it decreased under the error-level by the end of
the nineties. (The outliers around 1987 are due to the biggest crash
in the history of NYSE, called Black Monday when the market index
decreased by almost 20\%. As a consequence the motion of the returns
got highly synchronised \cite{jpo.blackmonday}.)
This result tells us that pulling effect between larger and smaller
stocks on the daily scale, i.e. the price of larger stocks pulling the
price of smaller ones, essentially vanished during the twenty years.

\subsection{Time-dependent correlations on high frequency scale}
\label{time-dependent.correlations.on.high.frequency.scale}
We made computations in order to analyse the time-dependent
cross-correlations of high frequency logarithmic returns and the
dynamics of these correlation functions. Our computations were carried
out in \textit{11} consecutive periods of the length of one
year. Using \eqref{eq:timedepcorr}, the window width in which the
logarithmic returns were calculated was \(\Delta t = 100\) seconds. We
altered \(\tau\) between \textit{-1000} seconds and \textit{1000}
seconds by steps of \textit{5} seconds. For the time shift between two
stocks the value of \textit{1000} seconds is beyond any reasonable
limit because of market efficiency, hence we also had the opportunity
to study the tail of the correlation functions, getting information
about their signal-to-noise ratio. As mentioned earlier, we confined
ourselves to analysing the 190 most frequently traded stocks in this
period.

\begin{figure}
\begin{center}
\psfrag{tau}[][][1.3][0]{\(\tau\) [sec]}
\psfrag{korr}[][][1.3][0]{\(C(\tau)\)} 
\hbox{
\includegraphics[width=54mm,height=70mm,angle=-90]{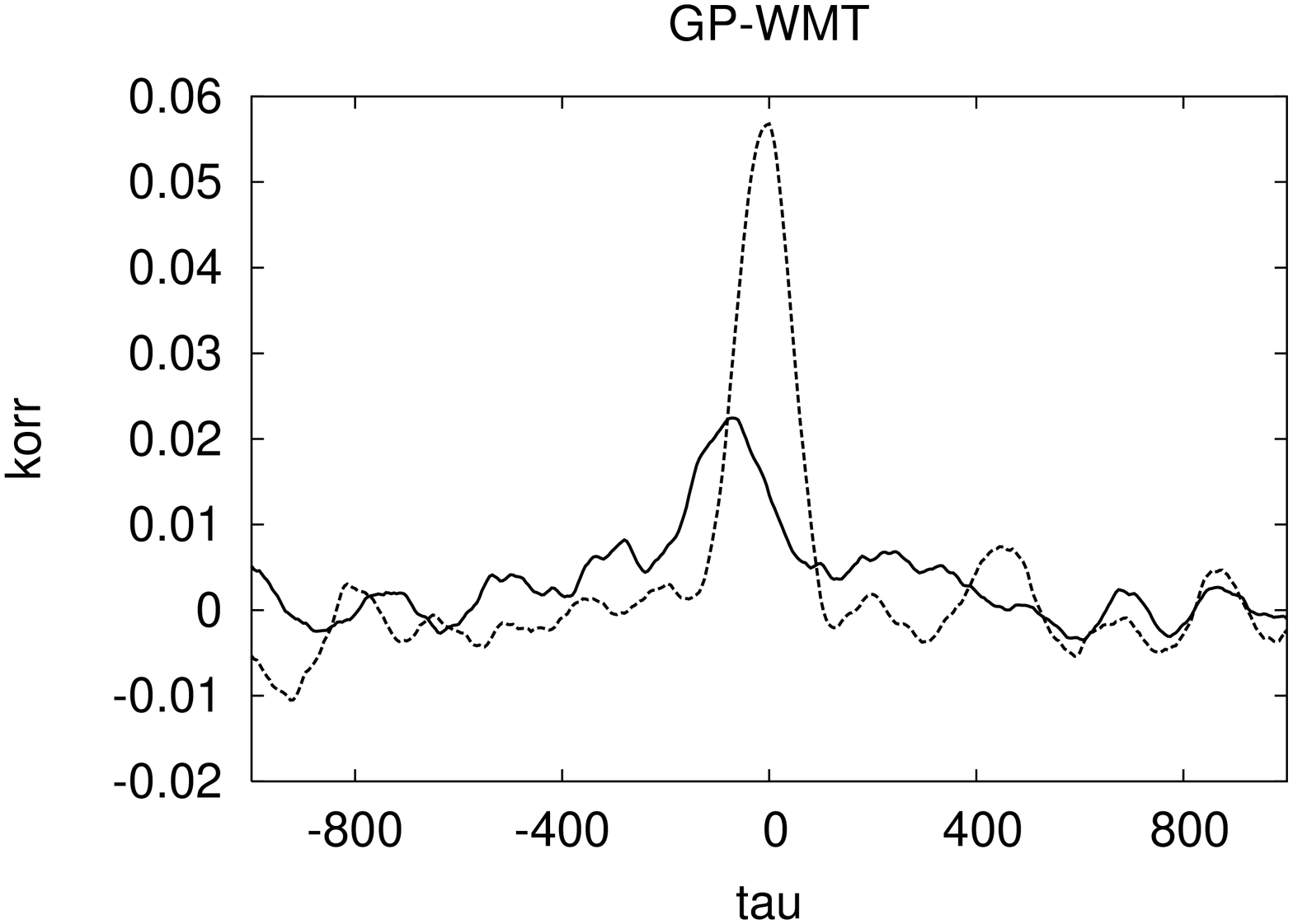}
\includegraphics[width=54mm,height=70mm,angle=-90]{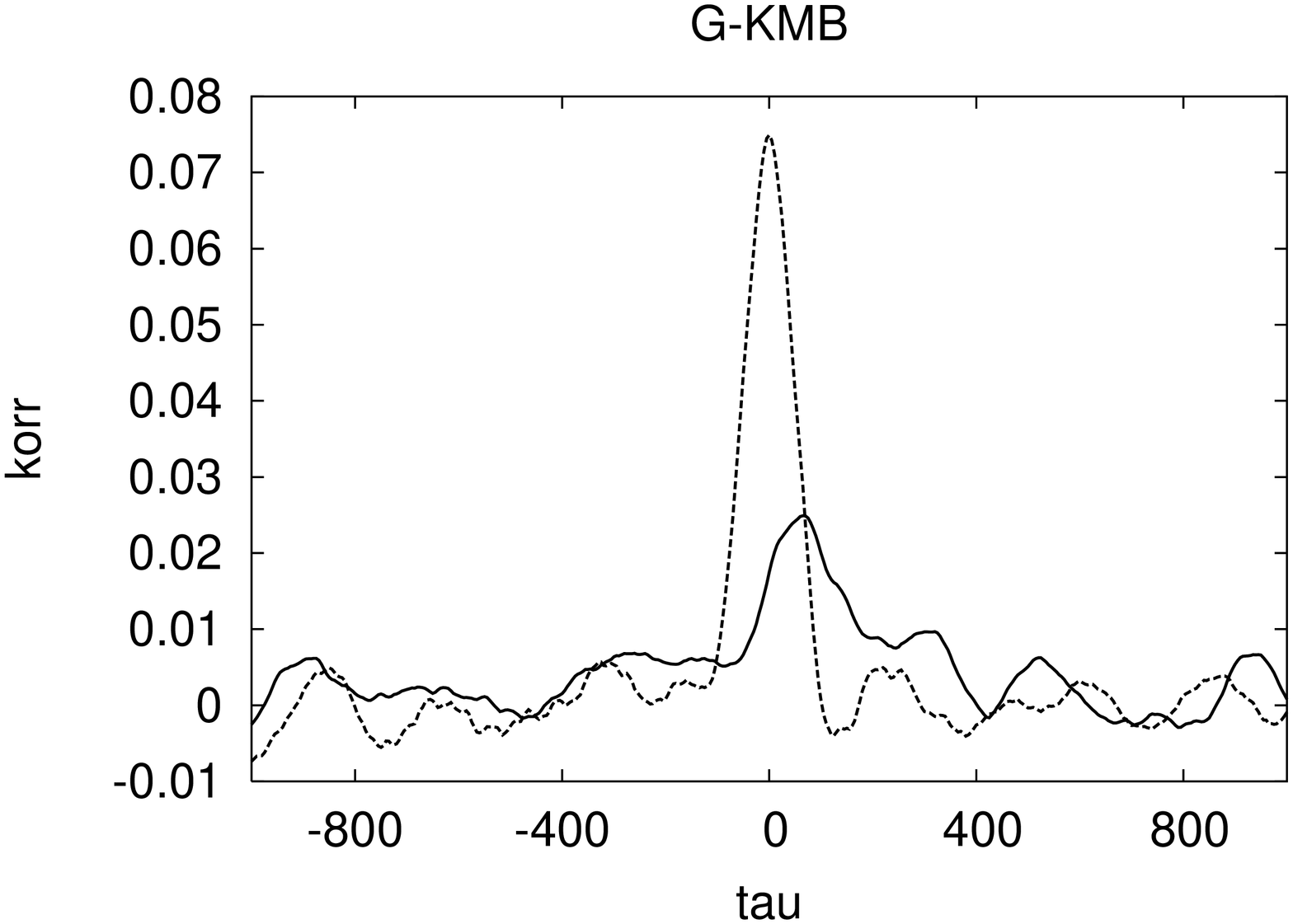} }

\hbox{
\includegraphics[width=54mm,height=70mm,angle=-90]{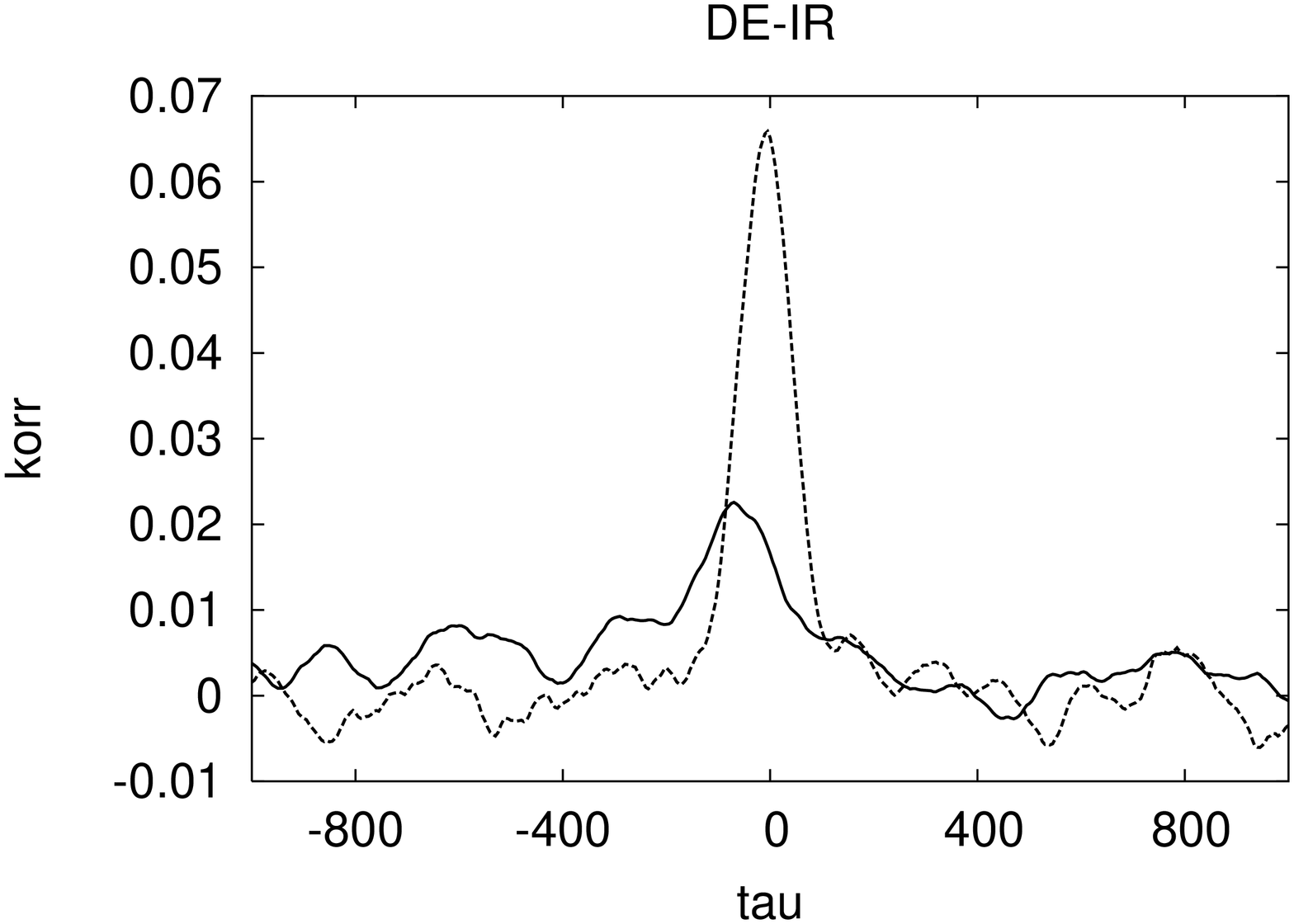}
\includegraphics[width=54mm,height=70mm,angle=-90]{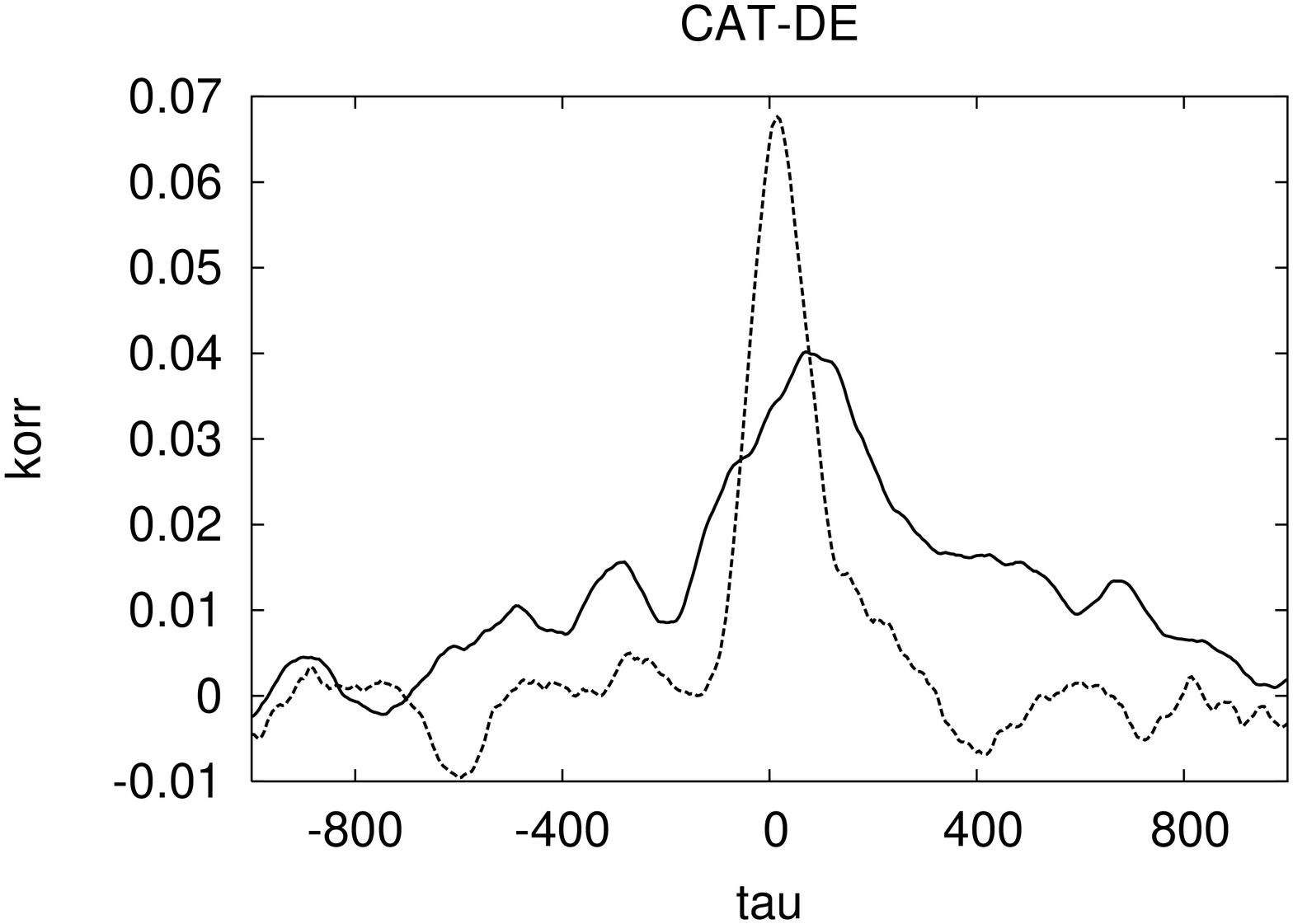}
}
\caption{Example plots showing the changes in the correlation
functions. The functions are plotted for the years 1993 (solid) and
2003 (dashed). The peaks have become much higher and the maximum
positions have moved towards zero.}
\label{example.plots1}
\end{center}
\end{figure}

\begin{figure}
\begin{center}
\psfrag{ev}[][][1.3][0]{year}
\psfrag{tau_ave}[][][1.3][0]{\(\overline{|\tau_{max}|}\) [sec]}
\includegraphics[height=100mm]{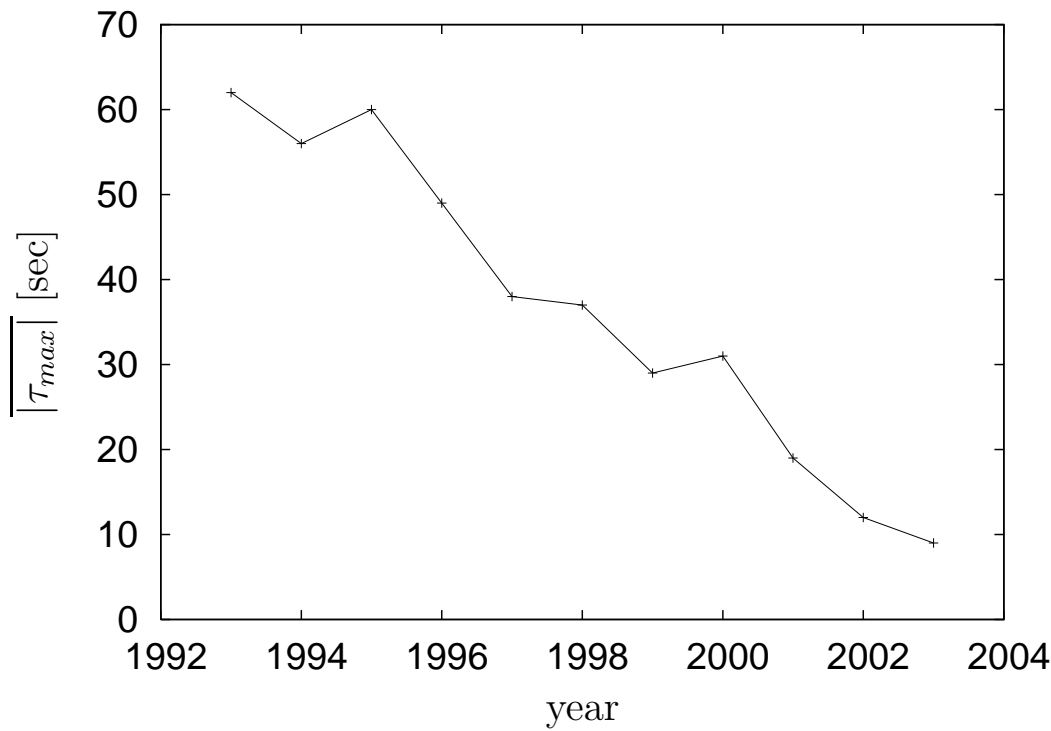}
\caption{The average time shift of the correlation functions taken
over every pair examined, as a function of time.}
\label{ave_tau}
\end{center}
\end{figure}

\begin{figure}
\begin{center}
\psfrag{tau2}[][][1.3][0]{\(\tau_{max}\) [sec]}
\psfrag{eloszlas2}[][][1.3][0]{\(P(\tau_{max})\)}
\psfrag{tau3}[][][0.8][0]{\(\tau_{max}\) [sec]}
\psfrag{eloszlas3}[][][0.8][0]{\(P(\tau_{max})\)}
\includegraphics[height=100mm]{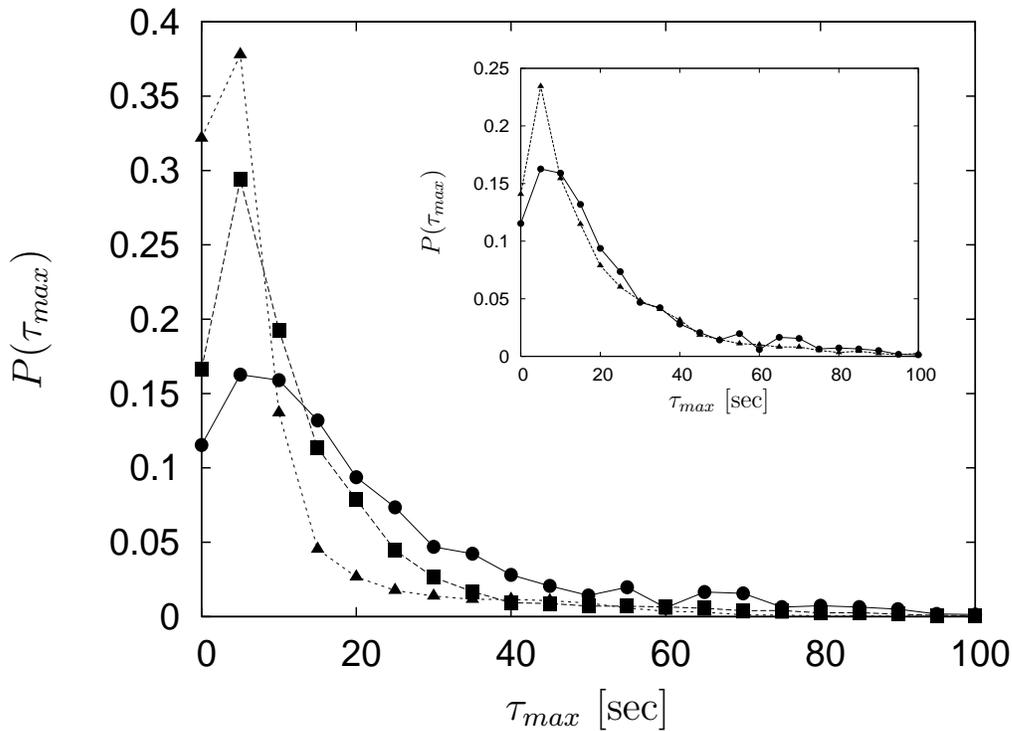}
\caption{Normalised distribution of the time shifts (\(\tau_{max}\))
of all time-dependent correlation functions having a maximum of
\(C_{max} > 0.02\) and \(\tau_{max} < 300\) seconds, for the years
1993 (circles), 1999 (squares) and 2003 (triangles). The distribution
becomes more and more sharply peaked near zero and thinner tailed
through the years. The inset shows the same distribution for the years
1993 (circles) and 2000 (triangles); we can see that the tails of the
two distributions are similar, indicating strong deviations from the
overall tendency.}
\label{distributions}
\end{center}
\end{figure}

We found that the time-dependent cross-correlation functions changed
significantly in the eleven years studied. Through the years the
maximum value (\(C_{max}\)) of the functions increased in almost all
cases and the time shifts (\(\tau_{max}\)) decreased. A few example
plots showing these effects can be seen on Figure
\ref{example.plots1}.
Figure \ref{ave_tau} shows the average of the absolute value of
maximum positions of the time-dependent cross-correlation functions
for every pair examined, as a function of the years. The decreasing
trend on the plot shows how the maximum positions approached the
ordinate axis through the years.
Figure \ref{distributions} shows the normalised distributions of the
maximum positions of all time-dependent correlation functions having a
maximum of \(C_{max} > 0.02\), in order to filter out those
correlation functions where no relevant peak can be found, and
\(\tau_{max} < 300\) seconds, in order to filter out peaks in the
correlation function due to the influence of two large logarithmic
return values instead of relevant lead-lag effects, for the years
1993, 1999 and 2003.  The distributions becoming more and more sharply
peaked near zero show the diminution of the time shifts. The change is
considerably strong, however, it is not monotonic. The inset in Figure
\ref{distributions} shows the same distributions for the years 1993
and 2000. We can see that the tails of the two distributions are very
similar indicating strong fluctuations from the tendency of the
changes in the time shift. Later we will see in the case of the equal
time correlation coefficients also, that 2000 is an outlier year, this
is possibly due to the dot-com crash.

The main cause of the changes in the correlation functions is the
acceleration of market processes. On the one hand computers and
various electronic equipments have spread in the last one or two
decades and their capacity and power have increased spectacularly,
resulting in largely increasing the speed of information flow and of
information processing.  On the other hand faster trading, shorter
periods between two transactions decrease the time of reaction of
market participants to information. These cause the decrease in the
time shift between the returns of two stocks and thus the maximum
position of the time-dependent cross-correlation function to move
towards zero. Furthermore, the decrease of the time shift, as well as
more synchronised reactions to the information result in growing
correlations of high frequency returns, i.e., in the diminution of the
Epps effect (see next section).

\subsection{Equal time correlations on high frequency scale}
\label{equal.time.correlations.on.high.frequency.scale}
\begin{figure}
\begin{center}
\psfrag{ev}[][][1.3][0]{year}
\psfrag{ave.rho}[][][1.3][0]{\(\overline{\rho}\)}
\includegraphics[height=100mm]{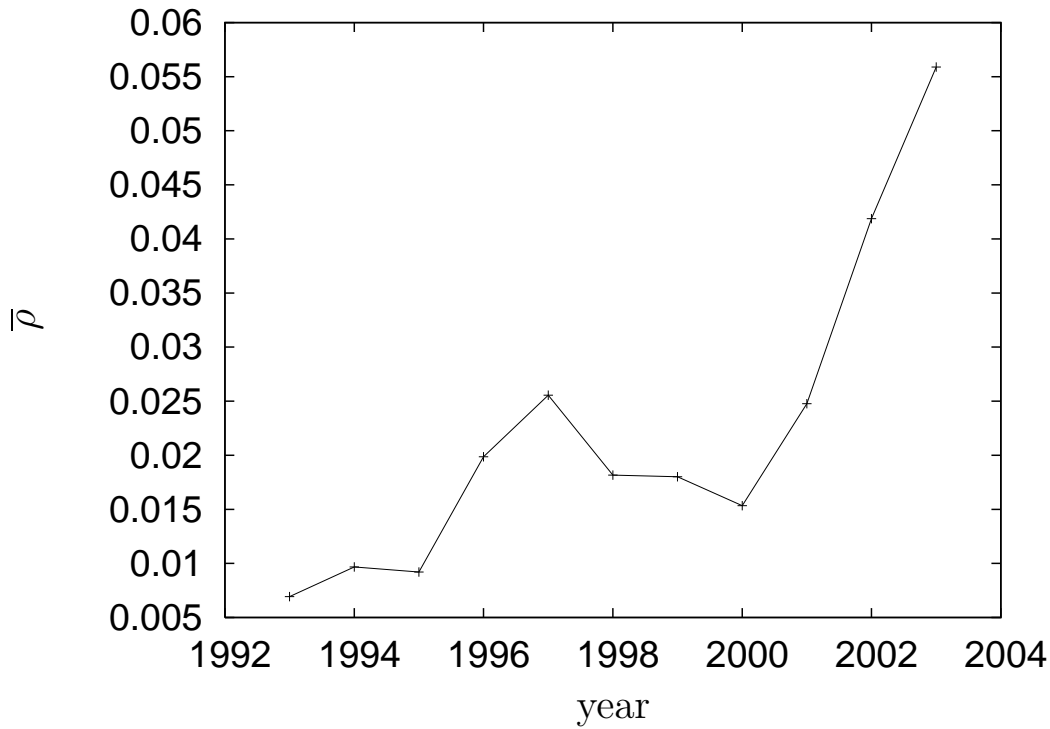}
\caption{The average of the equal time correlation coefficients of all
pairs examined as a function of time. There is a very strong though
not monotonic rise in the correlations.}
\label{ave_equaltime_corr}
\end {center}
\end{figure}

Since lagged correlations are possible causes of the Epps effect, i.e.
changes in equal time correlations, we also studied the dynamics of
equal time correlations as a function of time.  The computations were
carried out in \textit{11} consecutive periods of the length of one
year. Using \eqref{eq:corrcoefficient}, the window width in which the
logarithmic returns were calculated was \(\Delta t = 100\) seconds.
To see the general behaviour of the correlation coefficients, we
computed the equal weighted average of correlations for all pairs
examined. Figure \ref{ave_equaltime_corr} shows the average of the
correlation coefficients.  A very strong increase can be seen in the
correlation coefficients during the years. Nevertheless, the rise is
not monotonic, a local peak can be seen at the year 1997 and a local
minimum can be seen at the year 2000. We made separate computations in
case of closely and distantly related stocks, where the relations were
determined by the relative positions of the stocks on the minimum
spanning tree created for the period 1997--2000
\cite{mantegna.mst1,mantegna.mst2}. We considered two stocks being
close to each other if their distance was not greater than 3 steps on
the tree and being far from each other if their distance was not
smaller than 8 steps on the tree and for both the near related and
distant related pairs we computed the average correlation
coefficients. We found that the ratio of the two coefficients was
approximately constant until 1997, while after 2000 the correlation
coefficient of far laying stocks increased faster than that of close
ones. This differing change can be a sign of an equalisation process
of the correlations on the market.

The increase of the equal time correlations for high frequency data,
i.e. the diminution of the Epps effect can be traced back to two
reasons. One is the vanishing lagging as shown in the previous
paragraph.  Furthermore, the growing speed of market processes can be
understood as an expansion (lengthening) of the time scale of trades
on the stock market. Much more events, i.e. more averaging occur in a
certain length of time nowadays than did ten years ago. This higher
trading frequency acts against the Epps effect: The expansion of the
time scale brings larger correlations.

\section{Discussion}
\label{discussion}

We investigated the changes of the average pulling effect of weaker
stocks by stronger stocks on daily data of NYSE stocks. While for the
beginning of the eighties we have found an average correlation of
\textit{0.15--0.20} between the logarithmic returns of smaller stocks
and the previous days logarithmic returns of larger stocks, this
correlation decreased under the error level by the end of the
nineties. 
Since relevant time-dependent correlations on daily scale can be
exploited for arbitrage purposes, this finding is a sign of increasing
market efficiency. As trading and information procession get faster,
the time for each actor to react to the decisions of others'
decreases, so in order to exclude arbitrage opportunities (efficient
market hypothesis), time-dependent correlations on daily scale have to
diminish and vanish, and correlations -- if they exist -- must move to
a smaller scale (higher frequency).
This effect shows a considerable change in the structure of
the market, indicating growing market efficiency. 

We analysed time-dependent correlation functions computed between high
frequency logarithmic returns of NYSE stocks. We have found that the
positions of the peaks of the functions moved towards zero, and the
peaks got higher and sharper in the eleven years examined. The peak
approaching the ordinate axis is also a sign of growing market
efficiency. As trading got faster the reaction times and therefore the
time shifts decreased. Another consequence of faster reactions to
information is the diminution of the Epps effect, i.e. the equal time
correlations of high frequency returns increase with time. Not only
higher correlations but also sharper peaks are due to increasing
market efficiency.

We studied the dynamics of equal time cross-correlations of stock
returns on high frequency data . We have learnt that on the average
the correlations grew strongly, though the changes were not
monotonous. Correlations becoming larger indicate the diminution of
the Epps effect. This can be understood by the suppression of the
lagging in the correlations as well as by the fact that increasing
trading frequency causes an effective extension of the time scale,
enlarging the correlations.

The growing correlations and shorter time shifts in the time-dependent
cross-correlation functions are due to an increase of market
efficiency and diminution of the Epps effect. Since the origins of
these changes are present on all markets, they should also be possible
to find on markets different from the New York Stock Exchange.

\textbf{Acknowledgements}

Support of OTKA T49238 is acknowledged. Thanks are due to Z. Eisler,
J.-P. Onnela and G. Andor for their help.



\end{document}